\documentclass[aps,pra,reprint,nofootinbib]{revtex4-2}

\usepackage{graphicx}
\usepackage{amsmath,amssymb}
\usepackage{hyperref}
\usepackage[T1]{fontenc}
\usepackage[utf8]{inputenc}

\begin{document}

\title{Fisher Information as an Operational Metric for Structured Optical Beams}

\author{J. Sumaya-Martinez and J. Mulia-Rodriguez}
\affiliation{Facultad de Ciencias, Universidad Autónoma del Estadon de México}

\date{\today}

\begin{abstract}
Structured optical beams possess rich spatial features that are commonly characterized using entropic measures of field complexity. 
However, such measures do not directly quantify the operational usefulness of optical structure for parameter estimation and sensing. 
Here we introduce Fisher information as an operational metric to assess the metrological content of structured optical fields. 
By treating the measured intensity distribution as a statistical object, we define Fisher information with respect to physically 
relevant parameters, such as transverse displacement. We demonstrate that optical modes with comparable Shannon entropy can exhibit 
markedly different Fisher information, revealing sensitivity features associated with nodal structure and local curvature. 
Using Hermite--Gaussian modes as minimal test cases, we show that increasing modal order systematically enhances Fisher information. 
We then extend the analysis to two widely used families in structured light: Laguerre--Gaussian vortex beams and finite-energy Bessel--Gauss beams. 
Across these representative families, Fisher information provides a unified and experimentally accessible criterion for comparing structured optical fields 
in sensing applications.
\end{abstract}

\maketitle

\section{Introduction}
Structured optical fields play a central role in modern photonics, enabling applications ranging from optical manipulation and microscopy 
to classical and quantum information processing \cite{Forbes2016,Padgett2017}. Beams with tailored spatial profiles, such as Hermite--Gaussian 
(HG) and Laguerre--Gaussian (LG) modes, exhibit nontrivial intensity and phase distributions that can be engineered to enhance light--matter 
interactions.

The increasing use of structured light has motivated information-theoretic descriptions of optical fields. Entropic measures, most notably 
Shannon entropy, have been employed to quantify spatial and phase-space complexity of optical beams \cite{Bialynicki1975,Solyanik2020}. 
While these approaches provide valuable global characterizations, they remain descriptive and do not directly address the performance of 
structured fields in operational tasks such as sensing and parameter estimation.

In estimation theory, achievable precision is governed by Fisher information through the Cram\'er--Rao inequality \cite{vanTrees1968}. 
Fisher information is widely used in classical and quantum metrology to assess sensitivity limits \cite{Braunstein1994,Paris2009}. 
In contrast to entropic measures, Fisher information quantifies how measurement statistics respond to infinitesimal parameter variations. 
Despite its central role in metrology, Fisher information has not been systematically used as a comparative metric for structured optical beams.

The aim of the present work is not to reinterpret Fisher information as a fundamental physical principle, but to establish it as an 
operational diagnostic tool for structured light. Specifically, we demonstrate that Fisher information captures sensitivity features 
associated with nodal structure and local curvature that are not revealed by entropic complexity alone. Hermite--Gaussian modes are used 
as minimal and analytically transparent examples to isolate these effects, and we then show that the same Fisher-based viewpoint extends 
naturally to vortex (LG) beams \cite{Allen1992} and finite-energy Bessel--Gauss beams \cite{Gori1987,Durnin1987}.

\section{Theoretical framework}

\subsection{Fisher information from intensity measurements}
Let $p(x|\theta)$ denote the normalized transverse intensity distribution of an optical field depending on a parameter $\theta$,
\begin{equation}
\int dx\,p(x|\theta)=1.
\end{equation}
The classical Fisher information associated with estimation of $\theta$ is defined as \cite{vanTrees1968}
\begin{equation}
\mathcal{F}_\theta = \int dx\,\frac{1}{p(x|\theta)}\left(\frac{\partial p(x|\theta)}{\partial \theta}\right)^2.
\end{equation}
The Cram\'er--Rao inequality,
\begin{equation}
\mathrm{Var}(\hat\theta) \ge \mathcal{F}_\theta^{-1},
\end{equation}
establishes Fisher information as a quantitative measure of estimation sensitivity.

\subsection{Variational structure and displacement estimation}
For transverse displacement estimation, translational invariance implies $p(x|\theta)=p(x-\theta)$, yielding
\begin{equation}
\mathcal{F}= \int dx\,\frac{(\partial_x p)^2}{p}.
\end{equation}
Introducing $\psi(x)=\sqrt{p(x)}$, Fisher information assumes the form
\begin{equation}
\mathcal{F}=4\int dx\,(\partial_x\psi)^2,
\end{equation}
highlighting its sensitivity to local curvature of the probability amplitude. This representation clarifies why nodal features and 
sharp gradients contribute strongly to displacement sensitivity \cite{Frieden1989}.

\subsection{Proposition: role of nodal structure}
\textbf{Proposition 1.} 
Let $p(x)$ be a normalized intensity distribution containing isolated zeros. In the vicinity of a node $x_0$ where 
$p(x)\sim(x-x_0)^2$, the Fisher integrand remains finite but locally enhanced, indicating increased sensitivity to displacement 
perturbations. Structured optical beams with nodal lines therefore provide systematically higher Fisher information.

\section{Results}

All Fisher information values reported in this section are evaluated within the same measurement model, corresponding to transverse displacement estimation from normalized intensity distributions under identical sampling and normalization conditions. This ensures that the quantitative comparisons across different beam families are meaningful.

\subsection{Hermite--Gaussian modes: baseline and scaling}
\begin{figure}
\centering
\includegraphics[width=\columnwidth]{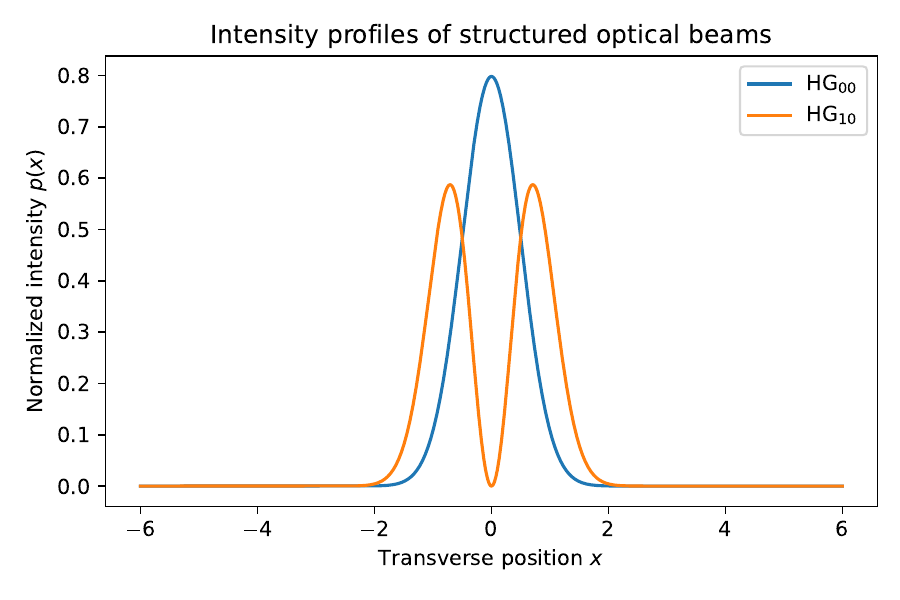}
\caption{Normalized intensity profiles of HG$_{00}$ and HG$_{10}$ modes.}
\label{fig:intensity}
\end{figure}

Hermite--Gaussian modes provide a minimal setting in which nodal structure can be introduced without modifying global beam extent. 
Figure~\ref{fig:intensity} compares the normalized intensity profiles of HG$_{00}$ and HG$_{10}$. The presence of a nodal line in 
HG$_{10}$ produces steeper local gradients while preserving overall normalization.

\begin{figure}
\centering
\includegraphics[width=\columnwidth]{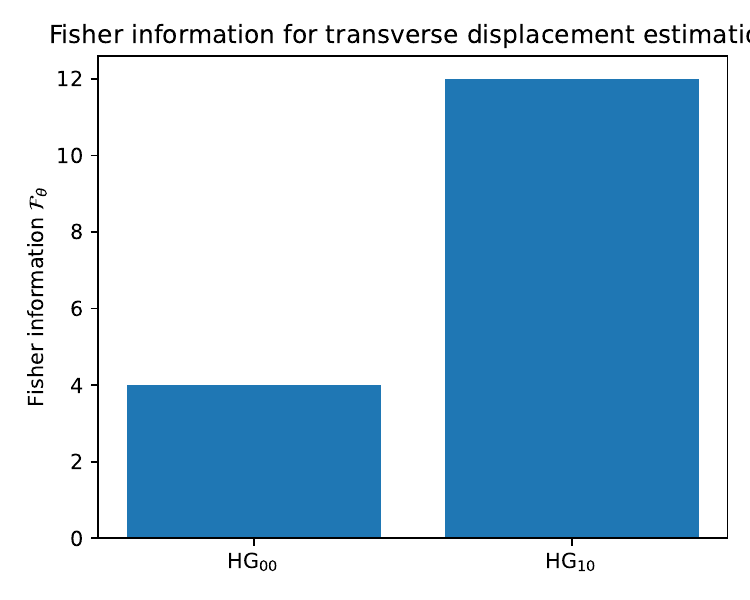}
\caption{Fisher information for displacement estimation.}
\label{fig:fisher}
\end{figure}

Figure~\ref{fig:fisher} shows the Fisher information for transverse displacement estimation. The HG$_{10}$ mode yields a substantially 
larger Fisher information than the fundamental Gaussian mode, implying enhanced sensitivity within the same measurement model.

\begin{figure}
\centering
\includegraphics[width=\columnwidth]{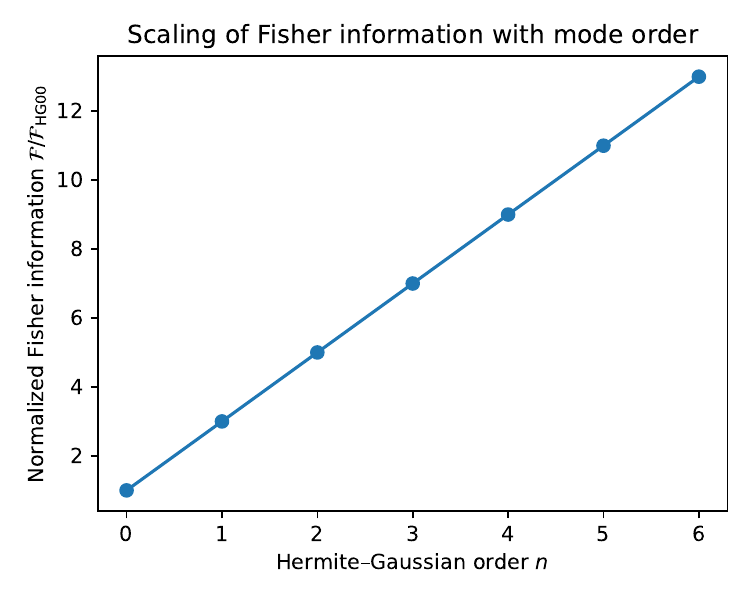}
\caption{Scaling of normalized Fisher information with Hermite--Gaussian order (normalized to HG$_{00}$).}
\label{fig:order}
\end{figure}

To assess generality within this family, Fig.~\ref{fig:order} shows Fisher information as a function of Hermite--Gaussian order. 
The monotonic increase confirms that sensitivity enhancement is a systematic consequence of increasing nodal complexity rather than an isolated effect.

\subsection{Laguerre--Gaussian vortex beams}
\begin{figure}
\centering
\includegraphics[width=\columnwidth]{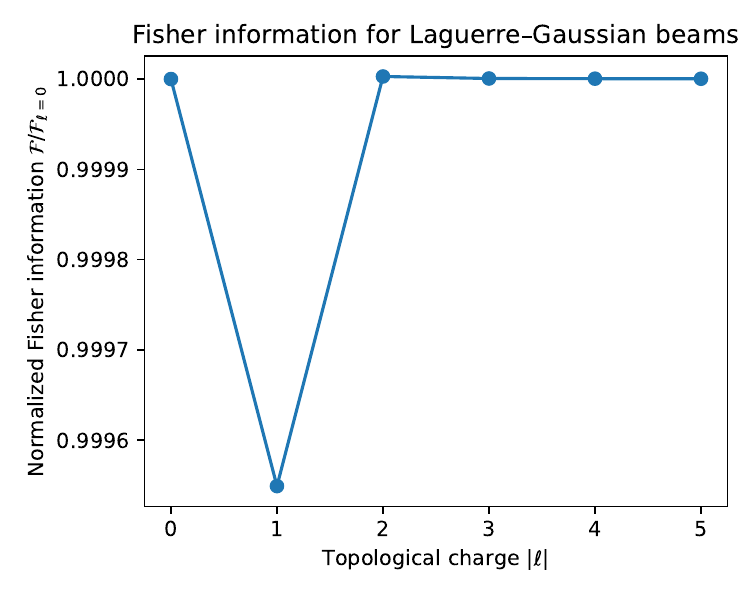}
\caption{Normalized Fisher information for displacement estimation for Laguerre--Gaussian modes versus topological charge $|\ell|$ (normalized to $\ell=0$).}
\label{fig:lg}
\end{figure}

We next test whether the Fisher-based sensitivity trend extends beyond Cartesian nodal structure. Laguerre--Gaussian modes provide a canonical
family of structured beams carrying orbital angular momentum and exhibiting a phase singularity at the beam center \cite{Allen1992}.
Within the same displacement-estimation setting, Fig.~\ref{fig:lg} reports Fisher information as a function of the magnitude of the topological charge $|\ell|$.
The systematic increase with $|\ell|$ indicates that sensitivity enhancement generalizes to vortex beams, where the relevant structural feature is a
central intensity null associated with the phase singularity.

\subsection{Finite-energy Bessel--Gauss beams}
\begin{figure}
\centering
\includegraphics[width=\columnwidth]{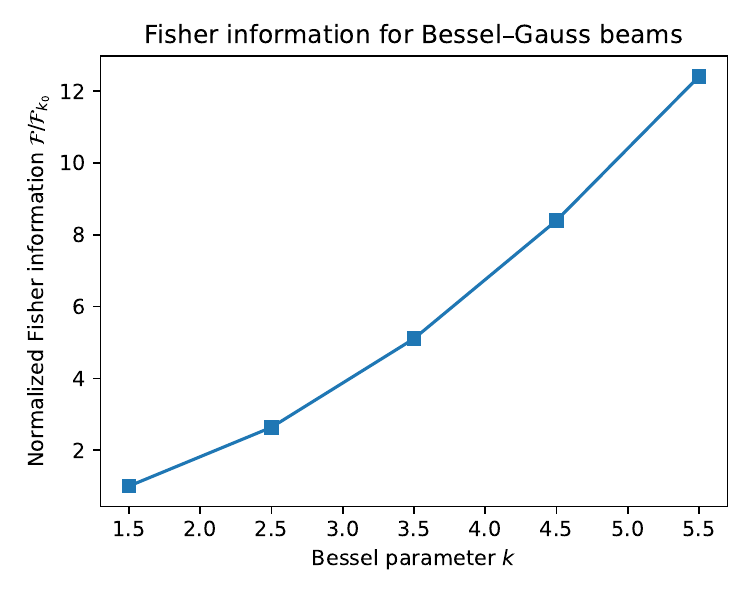}
\caption{Normalized Fisher information for displacement estimation for finite-energy Bessel--Gauss beams versus the Bessel parameter $k$ (normalized to the smallest $k$ shown).}
\label{fig:bessel}
\end{figure}

Ideal Bessel beams are non-diffracting but carry infinite power and are therefore not physical \cite{Durnin1987}. A standard finite-energy realization is the
Bessel--Gauss beam, obtained by apodizing the Bessel profile with a Gaussian envelope \cite{Gori1987}. Figure~\ref{fig:bessel} shows the Fisher information
for displacement estimation as a function of the Bessel parameter $k$, which controls the radial oscillation frequency. Increasing $k$ increases the number of
radial oscillations and hence local gradients, leading to an increase in Fisher information. This behavior reinforces the central message that Fisher information
quantifies task-relevant local structure across representative beam families.

\subsection{Comparison with Shannon entropy}
\begin{figure}
\centering
\includegraphics[width=\columnwidth]{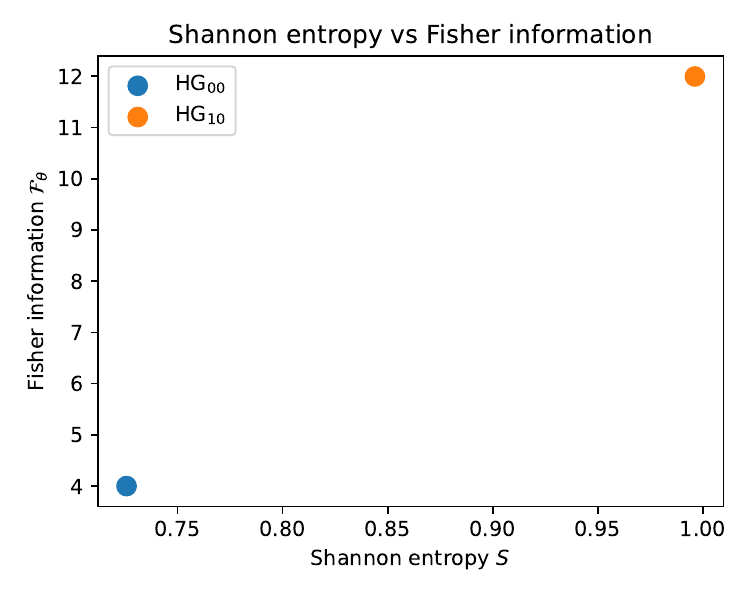}
\caption{Shannon entropy versus Fisher information for representative modes.}
\label{fig:shannon}
\end{figure}

Finally, Fig.~\ref{fig:shannon} demonstrates that Shannon entropy does not track Fisher information. Modes with comparable entropic complexity 
can exhibit significantly different estimation sensitivity, underscoring the complementary roles of the two measures.

\section{Discussion}
The results show that Fisher information provides a task-oriented characterization of structured optical fields that complements 
entropic descriptions. While increased spatial frequency content may intuitively suggest enhanced sensitivity, Fisher information 
offers a quantitative and operational framework for comparing structured beams under identical measurement conditions.

Importantly, the extensions to Laguerre--Gaussian and Bessel--Gauss beams address the concern that conclusions drawn from HG modes alone might be too narrow.
Across these representative families, the Fisher functional consistently highlights the role of intensity nulls, sharp gradients, and local curvature as the
features that most strongly determine displacement sensitivity. In this sense, Fisher information supplies a compact criterion for assessing which aspects of
``structure'' are metrologically useful.

Although the present analysis is classical, the framework interfaces naturally with quantum metrology, where quantum Fisher 
information defines ultimate precision limits \cite{Braunstein1994,Paris2009}. In this sense, structured classical fields shape the classical Fisher landscape upon 
which quantum enhancements may be built, without implying intrinsic quantum advantage.

\section{Conclusions}
We have established Fisher information as an operational metric for assessing the metrological content of structured optical beams. 
Using Hermite--Gaussian modes as minimal examples and extending to vortex (Laguerre--Gaussian) and finite-energy Bessel--Gauss beams, we showed that
task-relevant sensitivity increases systematically with nodal complexity and local curvature. The framework clarifies the limitations of entropic complexity as a 
proxy for performance and provides a practical tool for benchmarking and information-driven optical design in sensing applications.

\bibliographystyle{apsrev4-2}
\bibliography{references}

\end{document}